# The layered RuBr$_3$-RuI$_3$ honeycomb system

Danrui Ni, Xianghan Xu, Robert J. Cava*

Department of Chemistry, Princeton University, Princeton, NJ 08544, USA

*E-mail: rcava@princeton.edu

**Abstract**

The layered RuBr$_3$-RuI$_3$ honeycomb system was synthesized at high-pressures. The crystal structures are centrosymmetric (space group $R$-3), and based on honeycomb layers of Ru$^{3+}$ ($S$=1/2). Their basic physical properties are surveyed. The solid solution switches from insulating to metallic in the range of concentrations between RuBr$_{0.75}$I$_{2.25}$ and RuBr$_{0.50}$I$_{2.50}$. A preliminary structure/property phase diagram is presented. Our results suggest that this solid solution may provide insight into the influence of disorder on spin-orbit-coupled quantum spin liquids.

## 1.Introduction

Systems displaying magnetic frustration have attracted much attention in recent decades, as macroscopic degeneracies and qualitatively new states of matter can sometimes result, thus providing access to a wide range of behavior[1–4]. Among the studied systems, the honeycomb lattice with $S$=1/2 has stepped into the spotlight since the prediction of Kitaev quantum spin liquid states in such lattices[5], and honeycomb-layer-structured α-RuCl$_3$ has emerged as a candidate for Kitaev model physics[6–8]. In addition to α-RuCl$_3$, its sister compounds RuBr$_3$ and RuI$_3$, which crystallize in a 1D-chain structure at ambient pressure (similar to β-RuCl$_3$ [9]), also interest scientists and transfer to a honeycomb layered phase (the $α$-phase) under moderately high pressure[10–13]. With a similar layered structure, α-RuBr$_3$ and RuI$_3$ exhibit distinct physical properties, especially in that α-RuBr$_3$ is an insulator with an antiferromagnetic ordering transition, like α-RuCl$_3$, while α-RuI$_3$ is a metal with weakly paramagnetic behavior and no magnetic ordering down to 1.8 K[11,12,14,15]. This behavior inspired our exploration of the potential alloys between the Ru$^{3+}$-based tribromide and the Ru3+-based triiodide.

The Ru(Cl$_{1-x}$Br$_x$)$_3$ solid solution at ambient pressure crystallizes in a 1D-chain structure when the Br content is higher than in RuCl$_{2.4}$Br$_{0.6}$, and in a layer-related structure when the Br content is lower than 0.6[16]. High-pressure synthesis, however, enables the synthesis of a layered honeycomb solid solution between these materials over the full range of halide ratios. Based on what is reported, the honeycomb layered tribromide and the honeycomb-layered trichloride end-members have similar transport and magnetic properties[11,17,18], different from the layered

Ru tri-iodide[14]. Thus here we present the crystal structure and elementary properties of the honeycomb-layered α-Ru(Br$_{1-x}$I$_x$)$_3$ solid solution, with varying x, synthesized by a high pressure synthesis method. A honeycomb layered structure is found for each compound, and their physical properties reveal a strong variation with changing chemical composition, including an insulator-to-metal transition, a basic change in the temperature–dependent magnetic susceptibility, and the effect of introducing disorder on the anion site.

## 2. Experimental

Amorphous RuI$_3$ (Sigma-Aldrich, anhydrous, 99%) and crystalline RuBr$_3$ (Alfa Aesar, anhydrous, ⩾ 98%) were used as starting materials. The powders were mixed well in stoichiometric ratios and loaded into boron nitride crucibles. The mixtures were then inserted into a pyrophillite cube assembly, pressed to 6 GPa using a cubic multi-anvil system (Rockland Research Corporation), and heated to 800 °C at 50 °C/min with temperature determined by an internal thermocouple. The samples were kept at 800 °C for 3 hours and then quench-cooled before decompression. The products obtained were black and relatively stable in air. Small single crystals could be isolated in the post-reaction samples and were used for the single crystal X-ray diffraction characterization, while as-made dense pieces were saved for the magnetization, resistivity and heat capacity measurements.

A vapor transport method was employed on some of the post-synthesis samples to rid them of a small amount of chemical impurity, presumably introduced due to its presence in the starting materials. For this process, the samples were sealed in a quartz tube under vacuum with the hot end at 250 °C and the cold end at ambient temperature. The impurity was transferred to the cold end, and the honeycomb-layer structured solid solution samples were maintained in the hot zone with no signs of decomposition or phase transformation.

Single-crystal X-ray diffraction (SCXRD) was performed at 295 K using a Bruker D8 Quest Eco diffractometer equipped with a Photon III CPAD detector and monochromated Mo Kα radiation (λ= 0.71073Å). The refinement was performed by using the SHELXTL Software Package[19,20]. Powder X-ray diffraction (PXRD) characterizations were carried out on a Bruker D8 Advance Eco with Cu Kα radiation (λ = 1.5406 Å), and the Le Bail fitting of the acquired PXRD patterns was conducted via the TOPAS software.

Magnetization and heat capacity data were collected using a Quantum Design (QD) Dynacool PPMS-9, equipped with a vibrating sample magnetometer (VSM) option. Resistivity was also measured on QD Dynacool PPMS, using a four-probe method, and electrodes were made by silver epoxy. The magnetic field applied for the temperature-dependent magnetization measurements was 1000 Oe, and the magnetic susceptibility was defined as M/H where M is the measured Magnetization and H is the applied magnetic field. For the AC susceptibility measurements, a varying field of 4 Oe amplitude was applied with a DC bias field of 10 Oe.

## 3. Results and Discussion

The layered α-Ru(Br$_{1-x}$I$_x$)$_3$ honeycomb solid solution was synthesized by the high-pressure high-temperature method. The *R*-3 honeycomb layer structure was found by SCXRD for all compositions (**Figure 1A**) There is no sign of a missing center of symmetry, and bromine and iodine atoms appear to be totally disordered over the long range, occupying the same site in the unit cell. The crystallographic information and refined parameters of the solid solutions are listed in **Tables 1** and **S1-S2**. With increasing iodine ratio, the compound is revealed to have a higher density. A structural detail is noted, which is that for most of the refined structures, the normally empty interstitial sites (Ru2, Wyckoff position 3*a*) in this three-layer unit cell are occupied in the average structure by a small percentage of Ru atoms. This disordering of the Ru is most likely due to the presence of a small number of two-layer stacking faults in this three-layer structure, visible as a small number of Ru interstitials because diffraction experiments of the type performed are a positional average over the whole crystal. The presence of stacking faults like these is commonly observed in similar van der Waals layered-structure systems [12,21] so we are not surprised to see them here. The amount of Ru on the 3*a* site, and thus the fraction of two-layer stacking faults, is below 5% for almost all of the materials in the solid solution, while it is slightly higher in RuBr$_{0.5}$I$_{2.5}$ (where about 87% of the Ru occupies the honeycomb lattice while 13% occupies the interstitial site), which may suggest that there is a higher degree of interlayer stacking errors in RuBr$_{0.5}$I$_{2.5}$. Without constraints among the occupancy parameters in the structure refinement, these honeycomb materials freely refine to be slightly Ru-deficient (ranging roughly from Ru$_{0.92}$X$_3$ to RuX$_3$, where X = Br plus I). Thus we conclude that more detailed structural study, designed to observe stacking faults or other structural errors in this solid solution, such as by high-resolution electron microscopy, may be of future interest.

The RuX$_6$ octahedron of the solid solution series is close to ideal symmetry with a very slight distortion. Based on the bond angle and bond length data in **Table S2**, the Ru-X distance increases with increasing I in the formula, which is consistent with the expectation that iodine is larger than bromine. The PXRD characterization confirms the consistency of the bulk samples with the honeycomb structures observed by SCXRD, with the data for some representative compositions shown in **Figure 1B** and **Figure S1**. Similar diffraction patterns with shifted peak positions can be seen when comparing the PXRD patterns, indicating that they have the same structure type but varying lattice parameters. In the materials with higher bromine content, a small amount of layered RuBr$_3$ phase shows up as an impurity in the PXRD pattern (**Figure S1**). Compared to the end-member compounds, longer annealing time (3 hours) is required for the solid solutions to obtain a relatively uniform phase. With shorter annealing times (1 hour), lower diffraction intensity and wider diffraction peaks are observed in the products' PXRD patterns suggesting that a chemical composition distribution may be present if the samples are not heated long enough.

The magnetic susceptibility (M/H) was measured on polycrystalline samples at temperatures between 1.8 and 300 K (**Figure 2A**). The data for the end members, α-RuBr$_3$ and α-RuI$_3$, also synthesized in layered structures by the high-pressure method, are added as comparison.

Although all display a small upturn of magnetization at low temperature, typically attributed to a very small number of uncorrelated "orphan" spins[22], the materials in the α-Ru(Br,I)$_3$ solid solution can be divided into Br-rich and I-rich groups based on their magnetic behavior. The Br-rich compositions, whose magnetism is more strongly temperature-dependent, have a stronger magnetic response and clearer anomalies than the I-rich samples (with magnetism essentially temperature-independent). Among the solid solution materials, the material with composition RuBr$_2$I gives the highest magnetic susceptibility at low temperatures (its susceptibility at 100 K is above the general trend of the RuX$_3$ group, which will be presented later in the phase diagram) The 2:1 ratio of X ions is particularly interesting to us as it has the potential for short range ordering in this three-fold symmetry system.

In the Br-rich group, the temperature of the susceptibility anomaly (T$_A$) is around 32 K for α-RuBr$_3$, and 25-27 K for the solid solutions. This temperature range is labeled with different colors in **Figure 2A** for comparison. The decrease in temperature of the susceptibility anomaly may be interpreted as saying that by introducing iodine with disordered occupancy into the system, the frustration and randomness increase and thus suppress the ordering transition. To confirm that the magnetization anomalies observed in the layered honeycomb solid solution are not present due to the preserve of a small amount of the ambient pressure phase, an ambient pressure synthesis and magnetic characterization of 1D-chain structure "RuBr$_2$I" was also conducted (**Figure S2**), and no magnetic ordering down to 1.8 K was observed, indicating that the anomalies in **Figure 2A** can't arise from the presence of an adventitious 1D chain phase. In contrast, the I-rich group shows relatively weak paramagnetic behavior over a wide temperature range, especially for samples with x ≥ 0.75, which have no 3D magnetic ordering features visible down to 1.8 K.

While the I-rich samples show non-Curie-Weiss behavior in their magnetism even up to 300 K, Curie-Weiss fitting to the magnetic susceptibility in the high temperature range (150 – 300 K) for the Br-rich samples was carried out to further explore their magnetism (**Figure 2B** and **Figure S3**) using the following equation:

$$\chi - \chi_0 = \frac{C}{T - \theta} \qquad (1)$$

Some of the important fit parameters are listed in **Table 2**. The effective moment per Ru, (μ$_{eff}$), is larger than the theoretical value of a spin-only S=1/2 system (μ$_s$ = 1.73), confirming the expected non-negligible spin orbit coupling in the system. Additionally, based on the fitting results, RuBr$_{1.5}$I$_{1.5}$, the composition with the largest degree of Br-I mixing disorder, gives the largest frustration parameter. We thus deduce that the disorder induced by the off-magnetic-site mixing of nonmagnetic Br and I may affect the magnetic ground state of this honeycomb-based material by introducing disorder in the bond lengths and angles, which would change the energy states and orbital overlap. It is noted that the atypically large μ$_{eff}$ and θ values that these Curie-Weiss fittings find may suggest that the spin coupling in the α-Ru(Br,I)$_3$ system doesn't perfectly follow the standard Curie-Weiss law[14], and the fitting results here can be taken as an approximation. Because for a magnetic honeycomb lattice, relatively large frustration parameters and the deviation of zero-field-cooled (ZFC) and field-cooled (FC) curves suggest that the solid solution system may adopt spin-glass-like behavior[23,24], AC susceptibility measurements were conducted on an α-RuBr$_2$I sample. As presented in **Figure 2C**, these measurements show a variation with changing frequency, consistent with magnetic glassiness in the system. This is not surprising, as it is well established that disorder can lead to a spin-glass ground state. According to the theoretical literature, however, a frustrated system with both disorder and strong spin-orbit coupling can result in the emergence of different quantum-spin phenomena. Chemical disorder and the resulting randomization of magnetic interactions can enhance quantum fluctuations in a magnetic honeycomb system[25], suppress the long-range order, and lead to defect-induced frozen magnetic degrees of freedom[26]; inducing mimicry of a spin-liquid state in some materials[27]. Further exploration of the spin-orbit physics of this disordered layered honeycomb solid solution system may therefore be of future research interest to experts in that area of study.

The M vs H curves for different members of the solid solution were measured at 2 K and 250 K and are plotted in **Figure 3** and **Figure S4**. Same as the magnetic susceptibility, the α-Ru(Br,I)$_3$ solid solution can be roughly divided into Br-rich and I-rich groups based on their behavior. Samples show linear magnetization response with changing magnetic field at 250 K, as expected. However, at 2 K, a weak S-shape at lower fields can be observed for I-rich formulas (**Figure 3B**), while the Br-rich samples tend to show more linear response in M vs H, with a small hysteretic opening of the curve (**Figure 3C**). It is known that an S-shaped character will be displayed by materials where the available spins and the magnetization follow a Brillouin function relationship[28], which however is not the case for the α-Ru(Br,I)$_3$ solid solution. Our data therefore suggest that the RuBr$_3$-RuI$_3$ honeycomb solid solution seems to be far from an ideal paramagnetic system and that the Ru spins present in the materials cannot be considered as isolated[12].

Resistivity measurements were carried out to characterize the transport properties of the solid solution, and to determine the transition point from an insulator to a metal. In **Figure 4**, the normalized resistivity of the α-Ru(Br$_{1-x}$I$_x$)$_3$ series is plotted versus temperature from 125 to 275 K (to rule out the influence of surface moisture), and the inset shows the enlarged views of the iodine-rich group. It is clear that the alloyed compounds show insulating behavior at the higher Br contents, as their resistivities decrease with increasing temperature. But for x = 0.83 (RuBr$_{0.5}$I$_{2.5}$), the compound behaves as a bad metal. Generally, the resistivity decreases with increasing iodine ratio (except for the x = 0.5, RuBr$_{1.5}$I$_{1.5}$ sample), which is consistent with the transition from an insulator to metal; the crossover from insulating to metallic behavior is located between x = 0.75 and x = 0.83. For insulating formulas, the resistivity activation energy values are calculated based on the linear fitting of high temperature range selected from ln(ρ) vs 1/T plots, and are listed in **Table 2**. The decrease of the activation energy value with increasing iodine content supports a gradual insulator-to-metal transition in the system; a more obvious metallic component is seen for the iodine-rich group, especially for x ≥ 0.75, as the nonlinear shape of the ln(ρ) vs 1/T curve may suggest an unneglectable metallic contribution (**Figure S5**).

The heat capacity data were collected for materials in the honeycomb-layered α-Ru(Br,I)$_3$ solid solution between 2 and 150 K (**Figure 5A**). The phonon contribution to the fitting was estmated through the use of a modified Debye equation (Eq (2)[29]), with the total number of atoms in the formula fixed at 4, using a higher temperature range to avoid the potential magnetic contributions.

$$C_{phonon} = 9R \sum_{n=1}^{2} C_n \left(\frac{T}{\Theta_{Dn}}\right)^3 \int_0^{\Theta_{Dn}/T} \frac{x^4 e^x}{(e^x-1)^2} dx \qquad (2)$$

For RuBr$_{0.75}$I$_{2.25}$ and RuBr$_{0.5}$I$_{2.5}$, considering their metallic character based on the resistivity characterization, an electronic contribution $C_{electron} = \gamma T$ was also employed. γ is fitted to be 0.136 for RuBr$_{0.5}$I$_{2.5}$ and 0.083 for RuBr$_{0.75}$I$_{2.25}$, consistent with the expectation that the latter material is much less metallic than the former one.

By subtracting the phonon (and electron) contribution, a small magnetic contribution to the heat capacity ($C_{mag}$) can be estimated by looking at the lower temperature measurement range (examples are shown in **Figure S6**). The $C_{mag}$/T values obtained are plotted versus T in **Figure 5B** and the entropy released by the magnetic system on cooling (ΔS) can be calculated (**Figure 5B inset**). All of the solid solution compositions give a value obviously smaller than Rln(2$S$+1), suggesting that some amount of entropy has not been released down to 2 K, which may be due to the system frustration and randomness. Among these compositions, RuBr$_2$I is calculated to give the highest ΔS in the solid solutions, which is consistent with its stronger magnetic response. In contrast, RuBr$_{1.5}$I$_{1.5}$ gives the smallest value, which means that it holds most of the magnetic entropy to low temperature. This may be due to it having the highest degree of disorderly mixing and magnetic frustration, consistent with the behavior of its magnetic susceptibility.

No lambda-anomaly-like behavior is seen for any of the samples, suggesting that the anomalies observed at T$_A$ in the magnetic susceptibility may be due to the onset of short-range correlations instead of long-range order transitions. As a peak in the heat capacity of α-RuBr$_3$ was reported at around 34 K[11], our observations indicate that the disorder introduced by the Br/I mixing suppresses the long-range antiferromagnetic ordering of undoped α-RuBr$_3$. Similar behavior has been observed in a nickel oxyhalide system in 2022[30] where, with ordered occupancy of Cl, Sr$_2$NiO$_3$Cl presented low-dimensional antiferromagnetic behavior with a broad hump in the magnetic susceptibility, while isostructural Sr$_2$NiO$_3$F, where the fluorine anion occupies an apical site in a disordered manner with oxygen, exhibited spin-glass-like behavior with an anomaly at much lower temperature, and no anomaly associated with long-range magnetic ordering was observed in its heat capacity. The report demonstrated a long-range magnetically ordered state for anion-ordered Sr$_2$NiO$_3$Cl, and a short-range magnetically ordered state of anion-disordered $Sr_2NiO_3F$, with the same structure; an observation that is analogous to the disordered α-Ru(Br,I)$_3$ solid solution described here.

Summarizing the structural and physical properties characterized and discussed above, a phase diagram of α-Ru(Br$_{1-x}$I$_x$)$_3$ solid solution system can be drawn, and is shown in **Figure 6**. The lattice parameters *a* and *c* of the *R*-3 unit cell are plotted versus x in α-Ru(Br$_{1-x}$I$_x$)$_3$. Both parameters initially show relatively linear increase with increasing x but are then nonlinear when x is higher than 0.67, resulting a maximum of *c/a* at around x = 0.25 (*c/a*, a measure of the structural dimensionality, is frequently used to characterize layered materials); the solid solution system is more complicated than predicted by simple size-mixing expectations. Especially with a large degree of disorder by mixed occupancy on the same site, unconventional behavior can emerge. No matter what the details are, the phase diagram reveals a trend of the solid solution to evolve from a more RuBr$_3$-like behavior into a more RuI$_3$-like type behavior, as the alloyed compound changes, when x > 0.67, from that of a stronger magnetic response with short-range correlations (localized magnetism with Curie-Weiss-type behavior in the high temperature range measured), to a weakly paramagnetic behavior with no clear magnetic ordering (more and more delocalized and itinerant magnetism, with behavior not following the Curie-Weiss Law at high temperature) labeled as the orange-colored area, and changing from an insulator to a metal at the relatively high Iodine content of x = 0.83 (the shaded area).

## 4. Conclusion

The layered honeycomb α-Ru(Br$_{1-x}$I$_x$)$_3$ solid solution with varying x was prepared by a high pressure synthesis method. The *R*-3 unit cells of these materials, based on honeycomb-geometry Ru layers, were characterized by SCXRD, and their magnetic, transport properties, and low temperature heat capacities were determined. The variation with changing I to Br ratio was thus revealed, including an insulator-to-metal transition between RuBr$_{0.75}$I$_{2.25}$ and RuBr$_{0.5}$I$_{2.5}$, and a dramatic change in their magnetic properties. By introducing the disorder of halogen atoms, the frustration and randomness increased, triggering short-range magnetic correlations and what may be spin-glass-like behavior in the materials. This solid solution system may provide insight into the spin/orbit interaction in a spin ½ Ru-based system with disordered halogens. Further, more generally, it may provide a new venue for studying quantum spin phenomena with disorder and strong spin-orbit coupling for both magnetic frustrated insulators and metallic systems, and thus may open a door to further understanding and modifying the interaction of spin and orbital degrees of freedom of Kitaev systems. As it remains an open question how the ground states of the highly correlated, strongly fluctuating magnetic phases are affected by disorder, future more detailed study of the properties may be of interest. More generally, our results suggest that isovalent doping with anions from the same-column of the periodic table can trigger a complicated interplay among the electronic and magnetic states in honeycomb systems.

**Acknowledgement**

This research was funded by the Gordon and Betty Moore Foundation, EPiQS initiative, Grant No. GBMF-9066. X. Xu acknowledges the US Department of Energy grant DE-FG-02-98ER45706.

**Appendix**

The CCDC no. 2299103-2299108 contains the supplemental crystallographic data for this paper. These data can be obtained free of charge (available at www.ccdc.cam.ac.uk/data_request/cif, or by emailing data_request@ccdc.cam.ac.uk, or by contacting The Cambridge Crystallographic Data Centre, 12 Union Road, Cambridge CB2 1EZ, UK, fax: + 441223336033).

**Table 1.** Atomic coordinates and equivalent isotropic displacement parameters for α-Ru(Br,I)$_3$ at 300 K. ($U_{eq}$ is defined as one-third of the trace of the orthogonalized $U_{ij}$ tensor (Å$^2$)). The standard deviations are indicated by the values in parentheses.

Ru$_{0.924}$Br$_{0.5}$I$_{2.5}$:

| Atom | Wyck. | Occ. | x | y | z | $U_{eq}$ |
|---|---|---|---|---|---|---|
| I1 | 18f | 0.8333 | 0.33449 (9) | 0.32221 (10) | 0.58703 (3) | 0.0245 (3) |
| Br1 | 18f | 0.1667 | 0.33449 (9) | 0.32221 (10) | 0.58703 (3) | 0.0245 (3) |
| Ru1 | 6c | 0.805 (6) | 0.66667 | 0.33333 | 0.66677 (7) | 0.0177 (6) |
| Ru2 | 3a | 0.238 (9) | 0.33333 | 0.66667 | 0.66667 | 0.028 (3) |

Ru$_{1.009}$Br$_{0.75}$I$_{2.25}$:

| Atom | Wyck. | Occ. | x | y | z | $U_{eq}$ |
|---|---|---|---|---|---|---|
| I1 | 18f | 0.75 | 0.66524 (11) | 0.68219 (16) | 0.41270(4) | 0.0246 (8) |
| Br1 | 18f | 0.25 | 0.66524 (11) | 0.68219 (16) | 0.41270(4) | 0.0246 (8) |
| Ru1 | 6c | 0.9771 | 0.33333 | 0.66667 | 0.33304 (8) | 0.0242 (9) |
| Ru2 | 3a | 0.063 (17) | 0.66667 | 0.33333 | 0.33333 | 0.07 (3) |

Ru$_{0.923}$BrI$_2$:

| Atom | Wyck. | Occ. | x | y | z | $U_{eq}$ |
|---|---|---|---|---|---|---|
| I1 | 18f | 0.6667 | 0.6652 (2) | 0.6822 (2) | 0.41264 (6) | 0.0255 (5) |
| Br1 | 18f | 0.3333 | 0.6652 (2) | 0.6822 (2) | 0.41264 (6) | 0.0255 (5) |
| Ru1 | 6c | 0.906 (12) | 0.33333 | 0.66667 | 0.33311 (14) | 0.0201 (10) |
| Ru2 | 3a | 0.033 (14) | 0.66667 | 0.33333 | 0.33333 | 0.00 (3) |

Ru$_{0.937}$Br$_{1.5}$I$_{1.5}$:

| Atom | Wyck. | Occ. | x | y | z | $U_{eq}$ |
|---|---|---|---|---|---|---|
| I1 | 18f | 0.5 | 0.33458 (9) | 0.31968 (10) | 0.58770 (3) | 0.0283 (3) |
| Br1 | 18f | 0.5 | 0.33458 (9) | 0.31968 (10) | 0.58770 (3) | 0.0283 (3) |
| Ru1 | 6c | 0.896 (6) | 0.66667 | 0.33333 | 0.66686 (6) | 0.0203 (5) |
| Ru2 | 3a | 0.081 (7) | 0.33333 | 0.66667 | 0.66667 | 0.009 (6) |

Ru$_{0.957}$Br$_2$I:

| Atom | Wyck. | Occ. | x | y | z | $U_{eq}$ |
|---|---|---|---|---|---|---|
| I1 | 18f | 0.3333 | 0.68110 (12) | 0.66541 (10) | 0.58778 (4) | 0.0267 (3) |
| Br1 | 18f | 0.6667 | 0.68110 (12) | 0.66541 (10) | 0.58778 (4) | 0.0267 (3) |
| Ru1 | 6c | 0.951 (6) | 0.66667 | 0.33333 | 0.66698 (6) | 0.0196 (5) |
| Ru2 | 3a | 0.012 (3) | 0.33333 | 0.66667 | 0.66667 | 0.001 |

$Ru_{0.960}Br_{2.25}I_{0.75}$:

| Atom | Wyck. | Occ. | x | y | z | $U_{eq}$ |
|---|---|---|---|---|---|---|
| I1 | 18f | 0.25 | 0.6654 (5) | 0.6817 (6) | 0.41209 (16) | 0.0419 (10) |
| Br1 | 18f | 0.75 | 0.6654 (5) | 0.6817 (6) | 0.41209 (16) | 0.0419 (10) |
| Ru1 | 6c | 0.960 (17) | 0.33333 | 0.66667 | 0.3335 (4) | 0.0344 (17) |

**Table 2**. The magnetic anomaly temperature ($T_A$), Curie-Weiss temperature ($\theta$), frustration parameter ($f = |\theta| / T_A$), and the resistivity activation energy ($E_a$) calculated from linear fitting of the ln($\rho$) vs 1/T curves, for $\alpha$-Ru(Br$_{1-x}$I$_x$)$_3$. The activation energy value for x = 0 was obtained from Ref.11.

| Iodine Fraction x | $T_A$ (K) | $\theta$ (K) | f | $E_a$ (eV) |
|---|---|---|---|---|
| 0 | 32.3 | -145 | 4.5 | 0.21[11] |
| 0.25 | 27.1 | -157 | 5.8 | 0.066 |
| 0.33 | 25.8 | -322 | 12.5 | 0.054 |
| 0.5 | 25.6 | -564 | 22.0 | 0.049 |
| 0.67 | - | - | - | 0.007 |
| 0.75 | - | - | - | 0.004 |

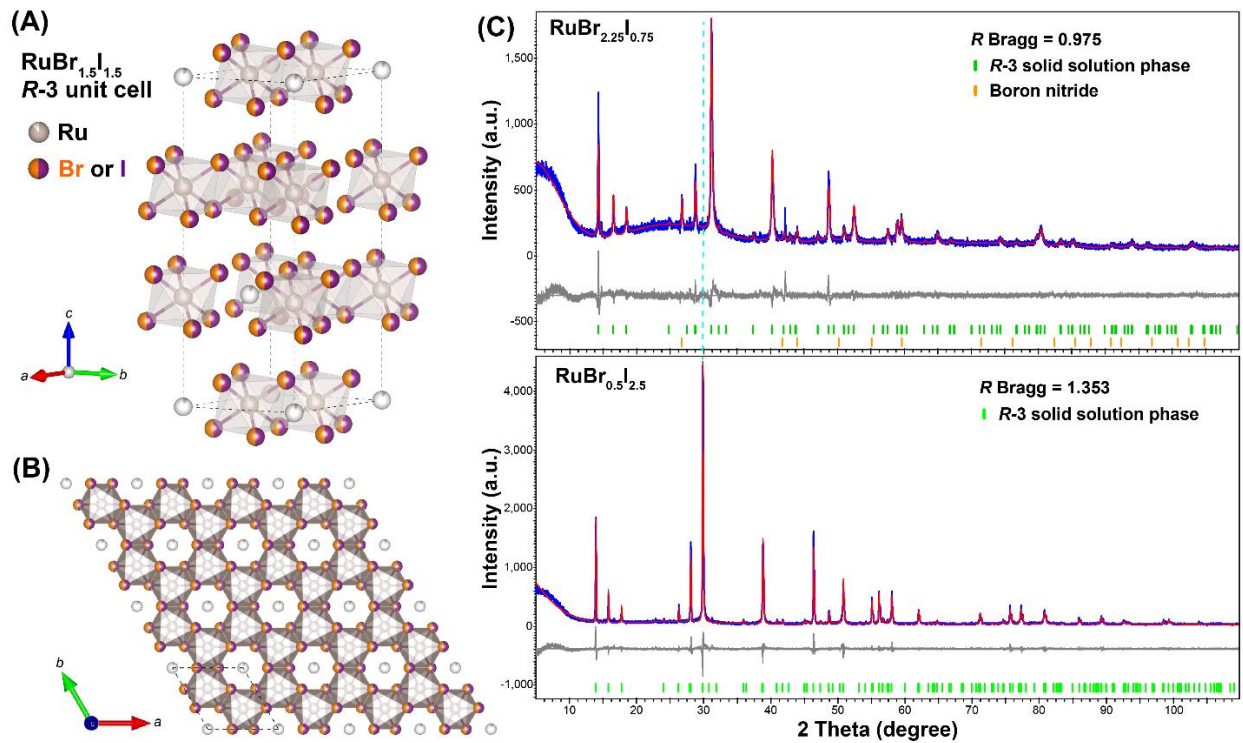

**Figure 1. (A)** The crystal structure of the honeycomb-layered α-RuBr$_3$-RuI$_3$ solid solution with a representative formula (x = 0.5 in Ru(Br$_{1-x}$I$_x$)$_3$ ) drawn, together with the view along the *c*-axis in **(B)**. **(C)** PXRD patterns with Le Bail fit for representative bulk Ru(Br$_{1-x}$I$_x$)$_3$ materials, confirming the consistency of the bulk sample structure and composition with the SCXRD refinement result. The dashed line shows the shift of the main peak position between the two patterns.

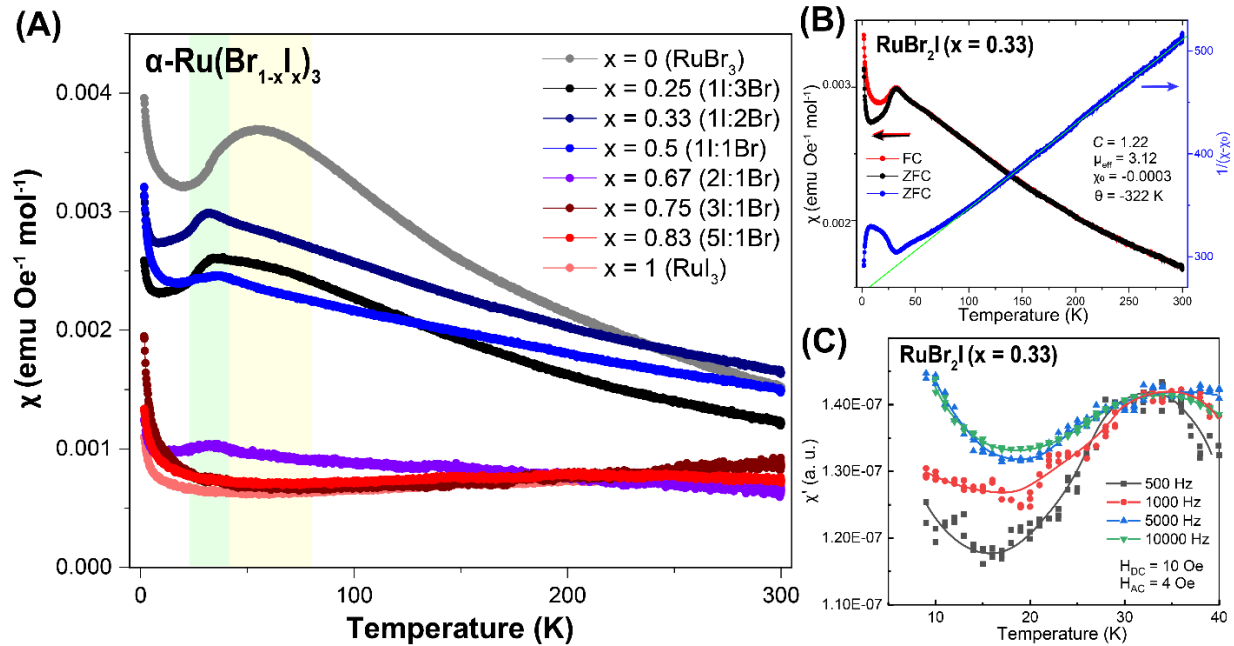

**Figure 2. (A)** The magnetic susceptibility measured from 1.8 to 300 K for the honeycomb-structure α-Ru(Br$_{1-x}$I$_x$)$_3$ solid solution series (0 ≤ x ≤ 1). Green colored shading marks the anomalies in the solid solution materials, while the yellow shading marks the anomaly in undoped α-RuBr$_3$. **(B)** Representative plot of the temperature dependent magnetic susceptibility χ (x = 0.33, ZFC in black and FC in red), together with the inverse of the difference between χ and the temperature-independent χ$_0$ in blue. Curie-Weiss fitting was conducted and the resulting parameters are shown in the panel. **(C)** The AC magnetic susceptibility data (χ′) collected on α-RuBr$_2$I (x = 0.33) from 10 to 40 K, at various frequencies under a small DC field (10 Oe) with an AC field (4 Oe) applied.

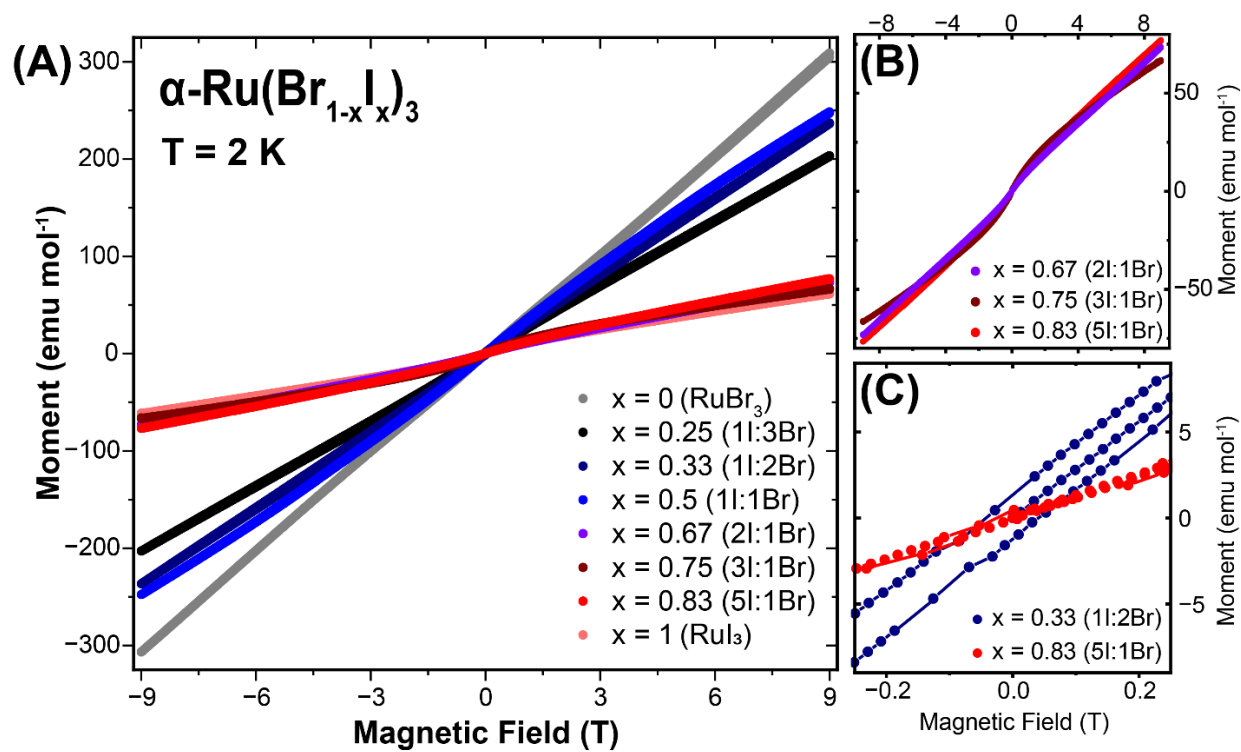

Figure 3. (A) The field-dependent magnetization data collected at 2 K for honeycomb-structure α-Ru(Br$_{1-x}$I$_x$)$_3$ (0 ≤ x ≤ 1), with the zoomed-in views of (B) the iodine-rich samples x = 0.67 to 0.83 to show the weak S-shape, and (C) x = 0.33 (Br-rich) and x = 0.83 (I-rich), revealing the very small hysteresis in this system.

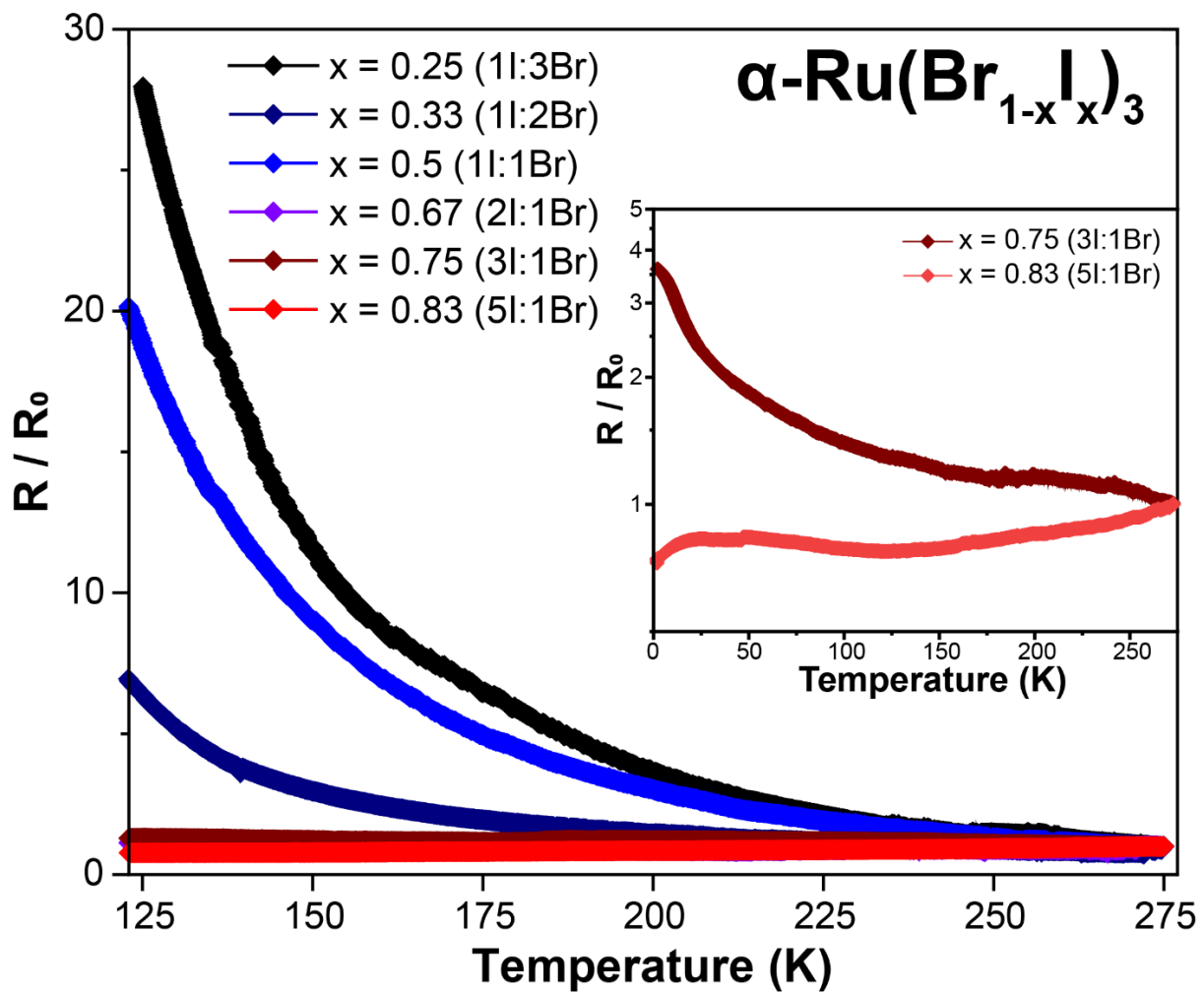

**Figure 4.** The normalized resistivity of the α-Ru(Br$_{1-x}$I$_x$)$_3$ series plotted versus temperature from 125 to 275 K, with the inset showing the enlarged views in the 1.8 to 275 K range for the iodine-rich group near the insulator-to-metal switching point (x = 0.75 and x = 0.83).

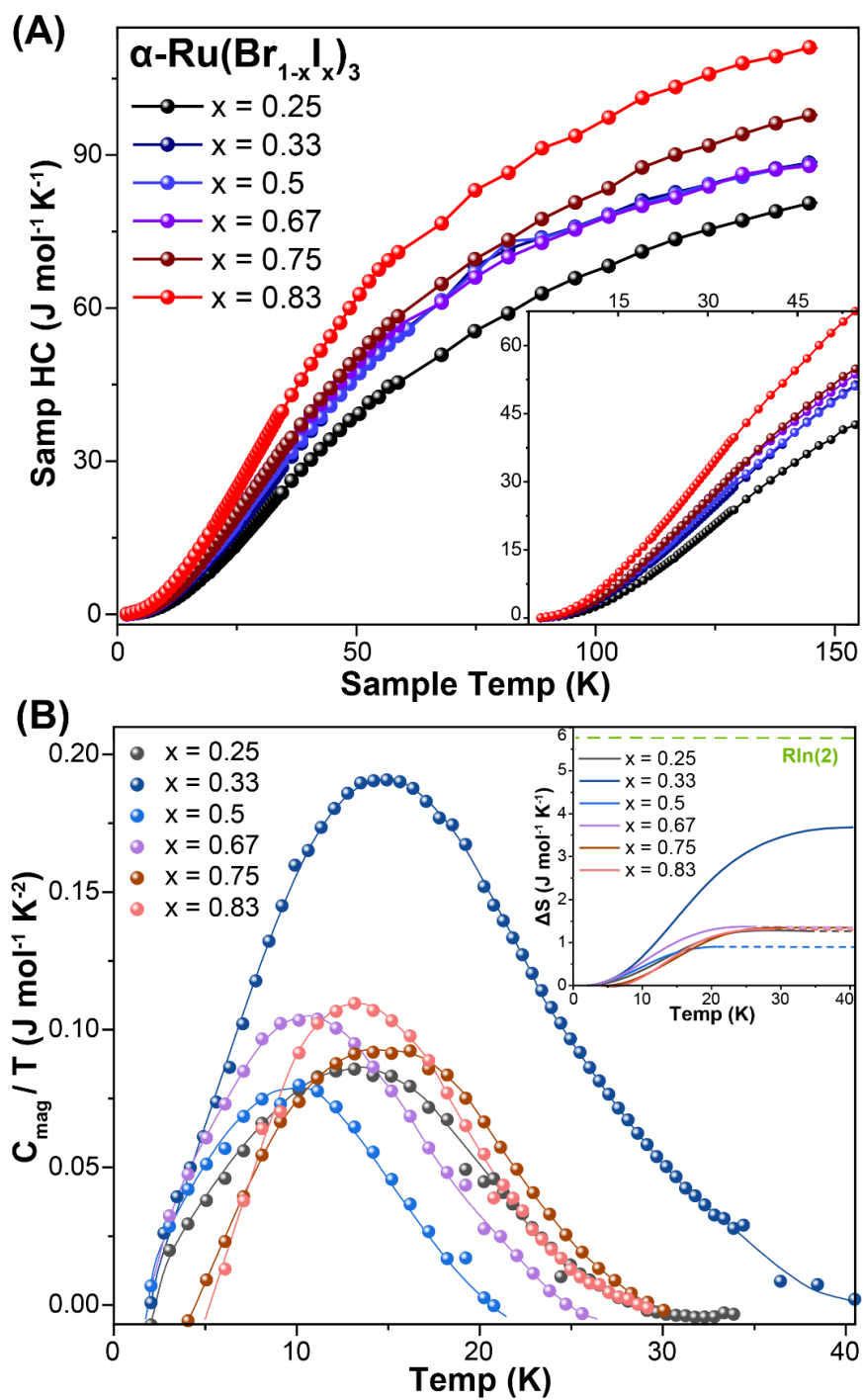

**Figure 5. (A)** The total heat capacity (HC) versus temperature of α-Ru(Br$_{1-x}$I$_x$)$_3$ from 2 to 150 K, with the 2 to 50 K range enlarged in the inset; **(B)** C$_{mag}$/T (the difference between the fit and the real data) plotted against temperature, with the corresponding ΔS calculated and plotted in the inset. The value of Rln(2) is labeled at the top of the inset as a reference.

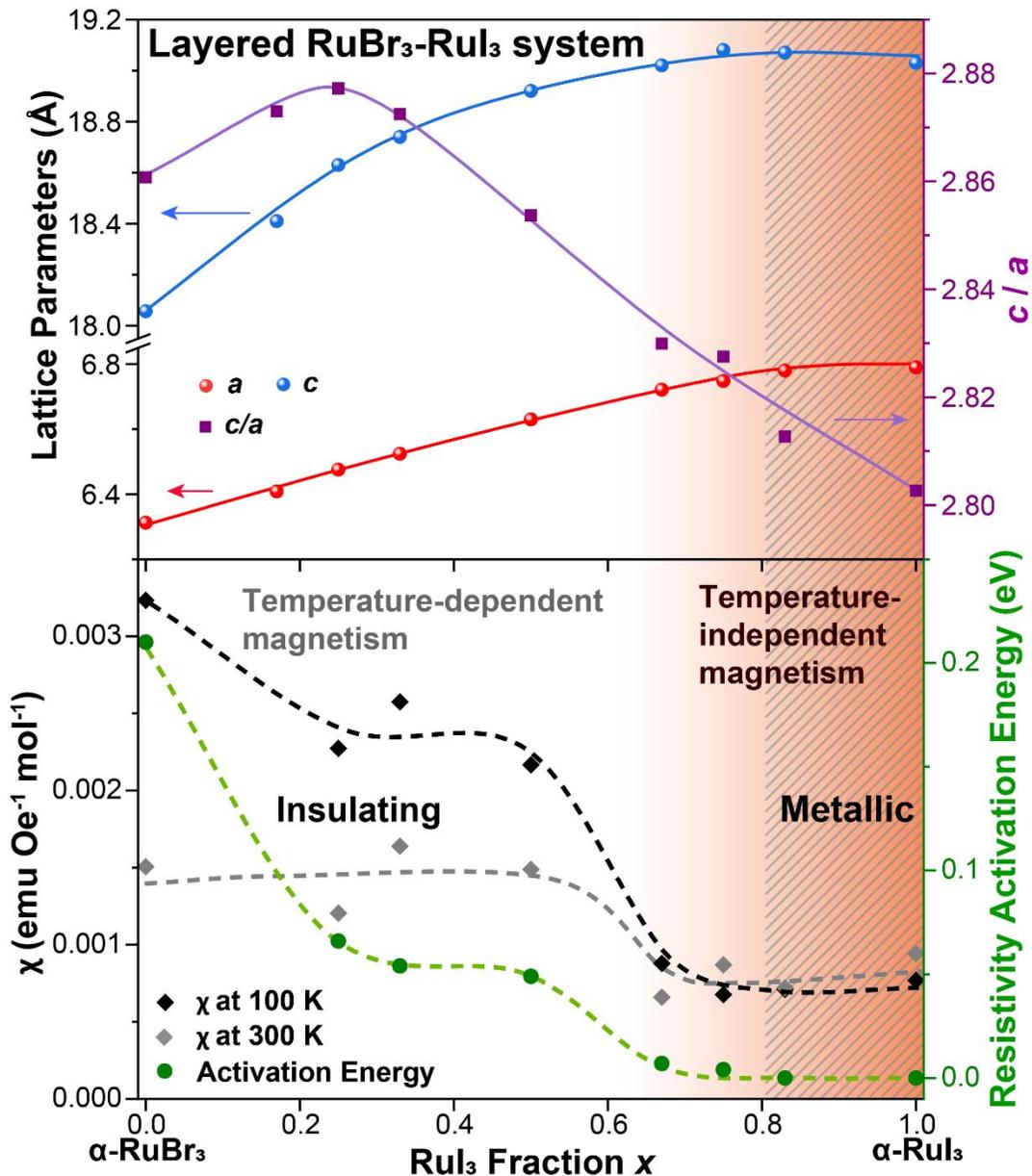

**Figure 6.** The tentative phase diagram for honeycomb-structure α-Ru(Br$_{1-x}$I$_x$)$_3$ (0 ≤ x ≤ 1). Lattice parameters *a* and *c* are plotted versus x, together with *c/a*, in the upper part; magnetic susceptibility χ at 100 K and 300 K, and the resistivity activation energy (Table 1) are plotted in the lower part. From the white to orange-colored areas the solid solution material changes from localized temperature-dependent magnetic into being weakly paramagnetic with no clear magnetic transition; from the white to the shaded area it also changes from insulating to metallic.

*Supporting Information for*

# The layered RuBr$_3$-RuI$_3$ honeycomb system

Danrui Ni, Xianghan Xu, Robert J. Cava*

Department of Chemistry, Princeton University, Princeton, NJ 08544, USA

*E-mail: rcava@princeton.edu

**Table S1.** Crystal Data and Structure Refinements for α-Ru(Br,I)$_3$ at 300 K. Standard deviations in temperature and cell parameters are indicated by the values in parentheses.

| Refined Formula | Ru$_{0.924}$Br$_{0.5}$I$_{2.5}$ | Ru$_{1.009}$Br$_{0.75}$I$_{2.25}$ | Ru$_{0.923}$BrI$_2$ | Ru$_{0.937}$Br$_{1.5}$I$_{1.5}$ | Ru$_{0.957}$Br$_2$I | Ru$_{0.960}$Br$_{2.25}$I$_{0.75}$ |
|---|---|---|---|---|---|---|
| Temperature (K) | 295(2) | 287(2) | 300(2) | 300(2) | 300 (2) | 300 (2) |
| F.W. (g/mol) | 450.69 | 446.52 | 426.90 | 404.34 | 383.44 | 371.99 |
| Crystal System | Trigonal | Trigonal | Trigonal | Trigonal | Trigonal | Trigonal |
| Space Group | $R$-3 (148) | $R$-3 (148) | $R$-3 (148) | $R$-3 (148) | $R$-3 (148) | $R$-3 (148) |
| Unit Cell Dimension $a$ (Å) | 6.7749(8) | 6.7413(11) | 6.7210(9) | 6.6379(6) | 6.5237(3) | 6.458(3) |
| Unit Cell Dimension $c$ (Å) | 18.990(3) | 18.993(4) | 19.026(4) | 18.947(3) | 18.7106(15) | 18.541(12) |
| Volume (Å$^3$) | 754.8(2) | 747.5(3) | 744.3(2) | 722.99(16) | 689.61(8) | 669.7(7) |
| Z | 6 | 6 | 6 | 6 | 6 | 6 |
| Calculated Density (g/cm$^3$) | 5.949 | 5.952 | 5.714 | 5.572 | 5.540 | 5.534 |
| Absorption Coefficient (mm$^{-1}$) | 22.012 | 22.902 | 23.229 | 24.884 | 27.157 | 28.481 |
| F(000) | 1144 | 1137 | 1089 | 1038 | 991 | 964 |
| Theta range for data collection | 3.218° to 24.722° | 3.218° to 24.695° | 3.212° to 24.692° | 3.226° to 25.359° | 3.266° to 24.501° | 3.296° to 24.744° |
| Index Ranges | -7 ≤ h ≤ 7, -7 ≤ k ≤ 7, -22 ≤ l ≤ 22 | -7 ≤ h ≤ 7, -7 ≤ k ≤ 7, -22 ≤ l ≤ 22 | -7 ≤ h ≤ 7, -7 ≤ k ≤ 7, -22 ≤ l ≤ 22 | -7 ≤ h ≤ 7, -7 ≤ k ≤ 7, -22 ≤ l ≤ 22 | -7 ≤ h ≤ 7, -7 ≤ k ≤ 7, -21 ≤ l ≤ 21 | -7 ≤ h ≤ 7, -7 ≤ k ≤ 7, -21 ≤ l ≤ 21 |
| Reflections Collected | 4480 | 4415 | 2733 | 4354 | 3356 | 2092 |
| Independent Reflections | 290 [R(int) = 0.0335] | 287 [R(int) = 0.0675] | 289 [R(int) = 0.0538] | 294 [R(int) = 0.0423] | 270 [R(int) = 0.0502] | 268 [R(int) = 0.1746] |
| Goodness-of-fit on F$^2$ | 1.204 | 1.152 | 1.144 | 1.152 | 1.164 | 1.131 |
| Final R Indices [I > σ(I)] | R1 = 0.0271, wR2 = 0.0636 | R1 = 0.0447, wR2 = 0.1241 | R1 = 0.0498, wR2 = 0.1103 | R1 = 0.0237, wR2 = 0.0604 | R1 = 0.0285, wR2 = 0.0646 | R1 = 0.1041, wR2 = 0.1310 |
| R Indices (all data) | R1 = 0.0307, wR2 = 0.0656 | R1 = 0.0491, wR2 = 0.1293 | R1 = 0.0591, wR2 = 0.1144 | R1 = 0.0367, wR2 = 0.0721 | R1 = 0.0411, wR2 = 0.0681 | R1 = 0.1654, wR2 = 0.1489 |
| Largest Diff. Peak and Hole (e.A$^{-3}$) | 1.300 and -1.190 | 2.026 and -2.155 | 2.456 and -1.222 | 1.063 and -1.636 | 1.200 and -1.403 | 1.534 and -1.912 |

**Table S2.** Selected bond lengths (Å) and bond angles (°) for α-Ru(Br,I)$_3$ at 300 K. X represents the atom on the 18$f$ site which is occupied in a disordered fashion by Br and I. Standard deviation is indicated by the values in parentheses.

|  | Ru$_{0.924}$Br$_{0.5}$I$_{2.5}$ | | Ru$_{1.009}$Br$_{0.75}$I$_{2.25}$ | | Ru$_{0.923}$BrI$_2$ | |
|---|---|---|---|---|---|---|
| **Bond Length (Å)** | Ru1-X (x3) | 2.6821(10) | Ru1-X (x3) | 2.6594(12) | Ru1-X (x3) | 2.6537(19) |
|  | Ru1-X (x3) | 2.6898(10) | Ru1-X (x3) | 2.6654(11) | Ru1-X (x3) | 2.6615(19) |
|  | Ru2-X (x6) | 2.7842(6) | Ru2-X (x6) | 2.7975(11) | Ru2-X (x6) | 2.7926(12) |
| **Bond Angle (°)** | X-Ru1-X (x3) | 91.55(4) | X-Ru1-X (x3) | 91.89(4) | X-Ru1-X (x3) | 92.04(5) |
|  | X-Ru1-X (x3) | 91.25(4) | X-Ru1-X (x3) | 91.35(5) | X-Ru1-X (x3) | 91.18(8) |
|  | X-Ru1-X (x3) | 90.73(3) | X-Ru1-X (x3) | 90.83(5) | X-Ru1-X (x3) | 90.70(8) |
|  | X-Ru1-X (x3) | 86.54(2) | X-Ru1-X (x3) | 86.07(3) | X-Ru1-X (x3) | 86.22(4) |
|  | X-Ru1-X (x3) | 177.06(2) | X-Ru1-X (x3) | 175.90(4) | X-Ru1-X (x2) | 175.90(5) |
|  |  |  |  |  | X-Ru1-X (x1) | 175.89(5) |
|  | Ru$_{0.937}$Br$_{1.5}$I$_{1.5}$ | | Ru$_{0.957}$Br$_2$I | | Ru$_{0.960}$Br$_{2.25}$I$_{0.75}$ | |
| **Bond Length (Å)** | Ru1-X (x3) | 2.6301(8) | Ru1-X (x3) | 2.5872(9) | Ru1-X (x3) | 2.554(5) |
|  | Ru1-X (x3) | 2.6363(8) | Ru1-X (x3) | 2.5910(9) | Ru1-X (x3) | 2.568(5) |
|  | Ru2-X (x6) | 2.7501(6) | Ru2-X (x6) | 2.7101(7) |  |  |
| **Bond Angle (°)** | X-Ru1-X (x3) | 91.70(3) | X-Ru1-X (x3) | 92.02(3) | X-Ru1-X (x3) | 92.14(14) |
|  | X-Ru1-X (x3) | 91.10(3) | X-Ru1-X (x3) | 90.97(3) | X-Ru1-X (x6) | 90.7(2) |
|  | X-Ru1-X (x3) | 90.70(3) | X-Ru1-X (x3) | 90.46(4) | X-Ru1-X (x3) | 86.57(10) |
|  | X-Ru1-X (x3) | 86.61(2) | X-Ru1-X (x3) | 86.67(2) | X-Ru1-X (x2) | 176.09(13) |
|  | X-Ru1-X (x3) | 176.42(2) | X-Ru1-X (x2) | 176.21(3) | X-Ru1-X (x1) | 176.08(13) |
|  |  |  | X-Ru1-X (x1) | 176.22(3) |  |  |

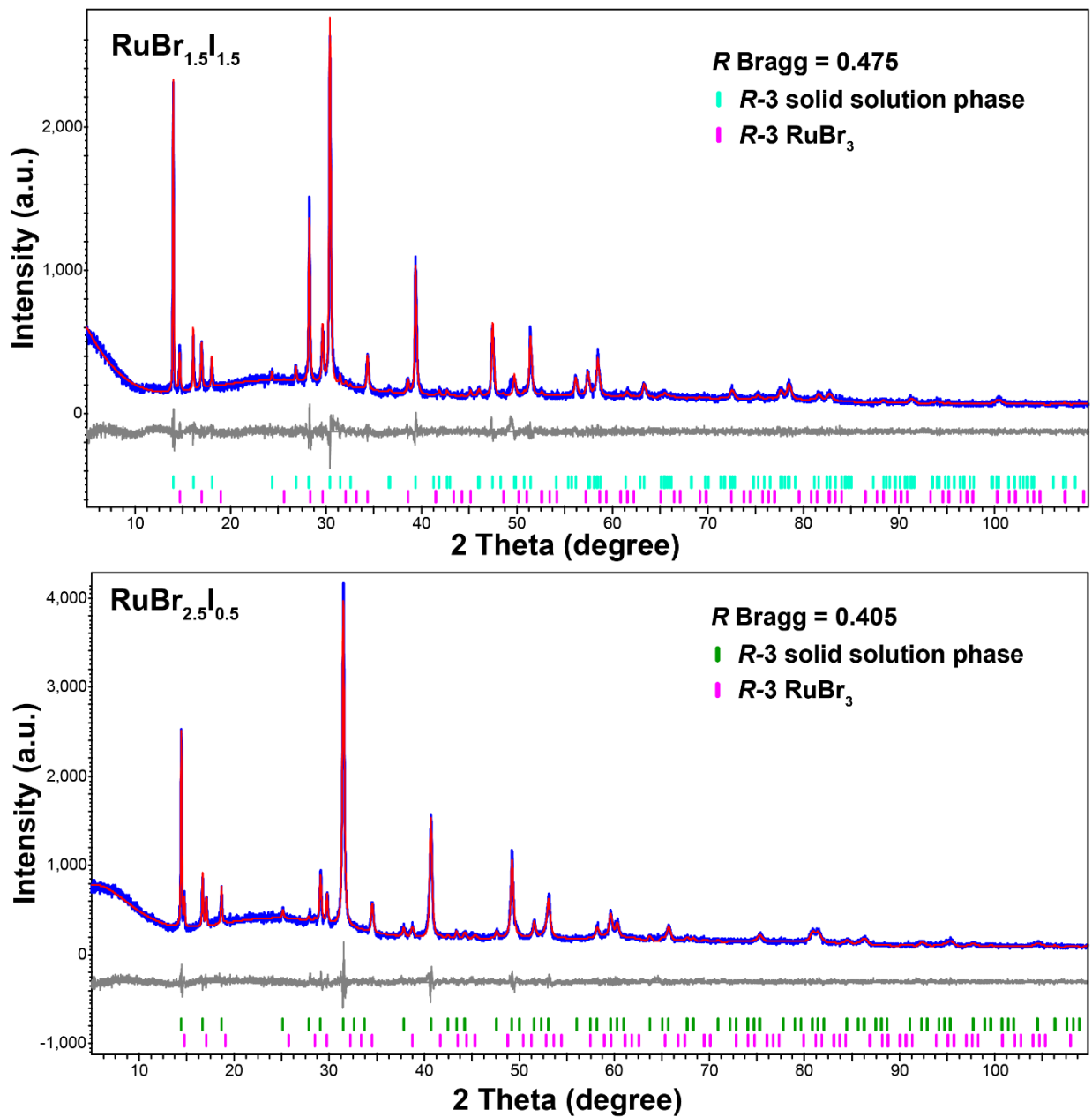

**Figure S1.** Representative PXRD patterns with Le Bail fit for bulk α-Ru($Br_{1-x}I_x$)$_3$ materials (top x = 0.5, bottom x = 0.17), confirming the consistency of the bulk sample structure and composition with the SCXRD refinement results.

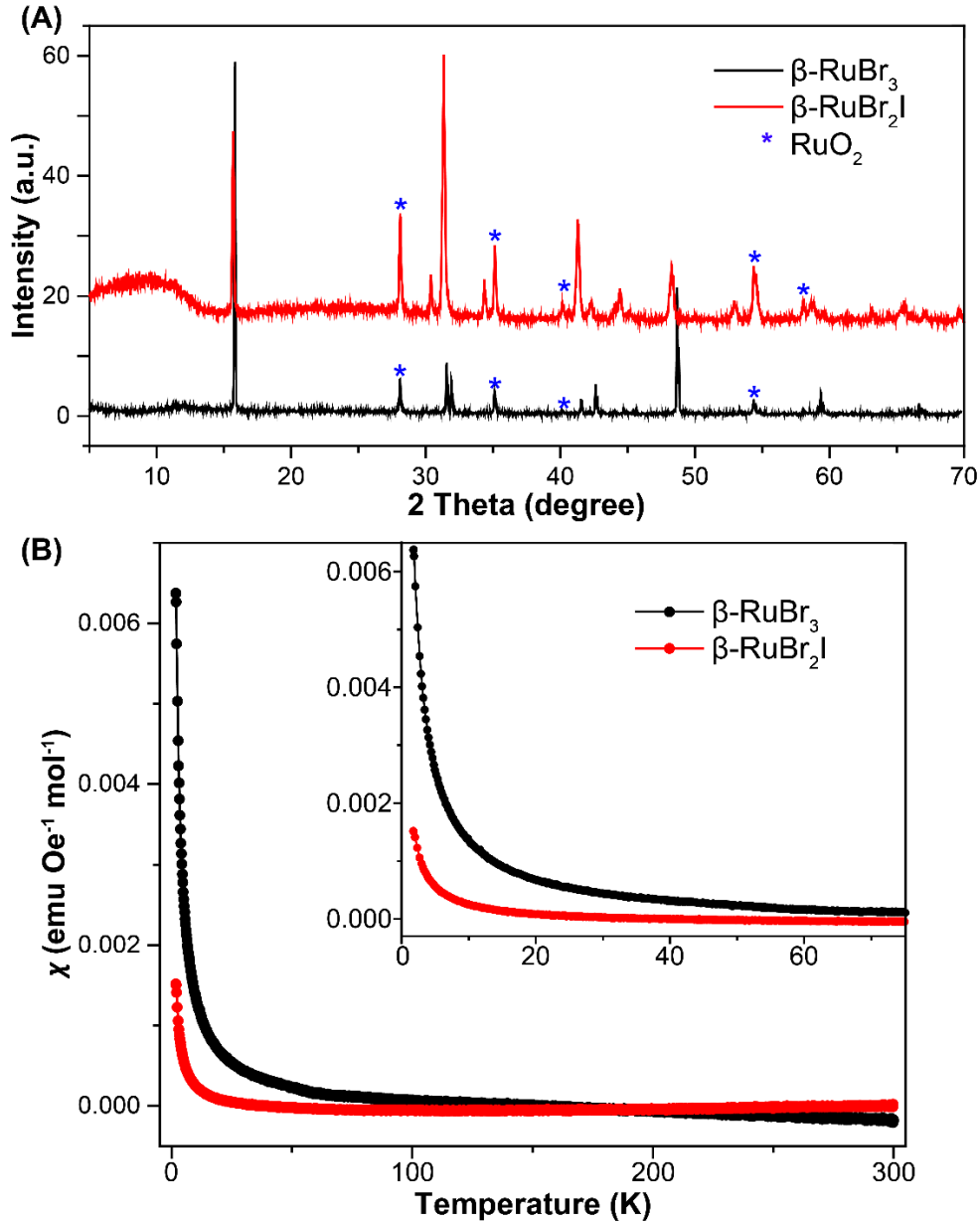

**Figure 2S.** (A) PXRD patterns of undoped 1D-chain structure β-RuBr$_3$ (black pattern), and isostructural β-RuBr$_2$I (red pattern) synthesized at ambient pressure (annealed at 400 – 600 °C under vacuum in sealed quartz tube for five days). RuO$_2$ impurity phase is labeled with * in the pattern. Comparing the two patterns, the shift of peaks with similar diffraction type suggests the iodine successfully doped into the 1D chain lattice without changing the structure. (B) Temperature dependent magnetic susceptibility (χ) of β-RuBr$_3$ and β-RuBr$_2$I, measured under 1000 Oe (ZFC) from 1.8 to 300 K. The inset shows a zoomed-in view of the 1.8 to 75 K range, confirming that no 3D magnetic ordering or transitions can be observed in the low temperature range for both samples.

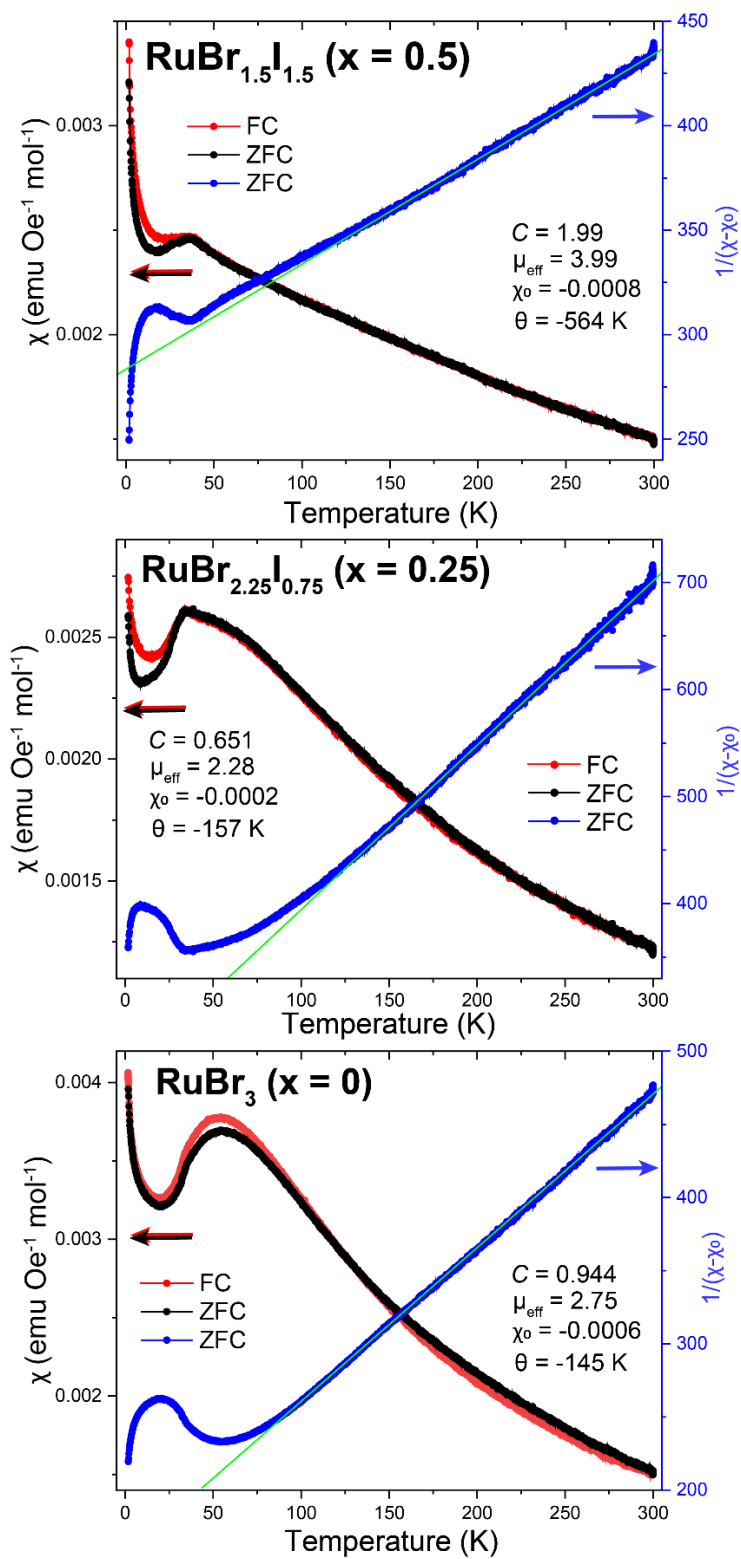

**Figure S3.** The χ vs. T plots of honeycomb-structure α-Ru(Br$_{1-x}$I$_x$)$_3$ (ZFC in black and FC in red), with the 1/( χ - χ$_0$) vs. T curves in blue color. Curie-Weiss fitting is conducted and the parameters are labeled in the panel. Top x = 0.5, middle x = 0.25, bottom x = 0.

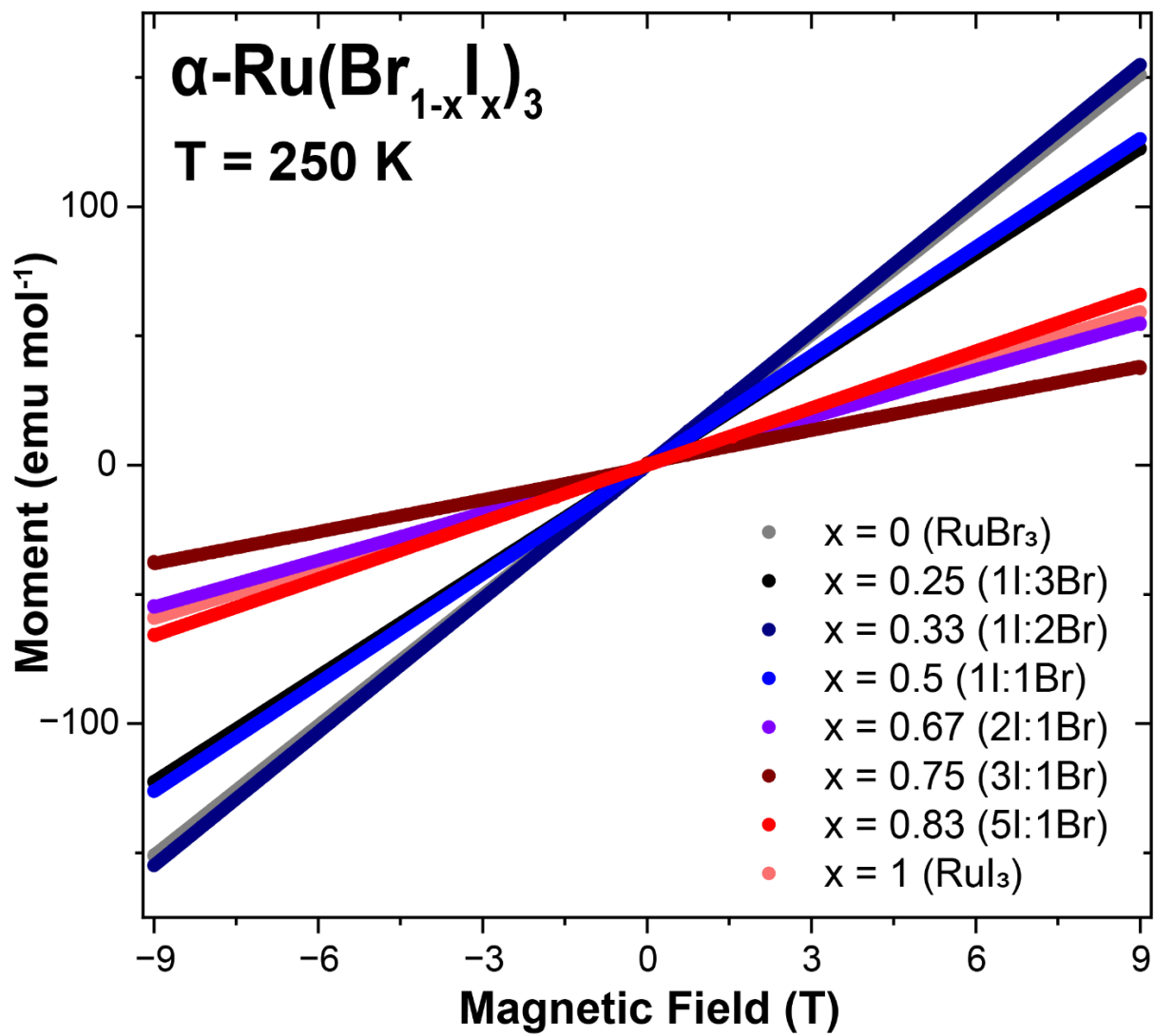

**Figure S4.** The field-dependent magnetization data collected at 250 K of honeycomb-structure α-Ru(Br$_{1-x}$I$_x$)$_3$ (0 ≤ x ≤ 1).

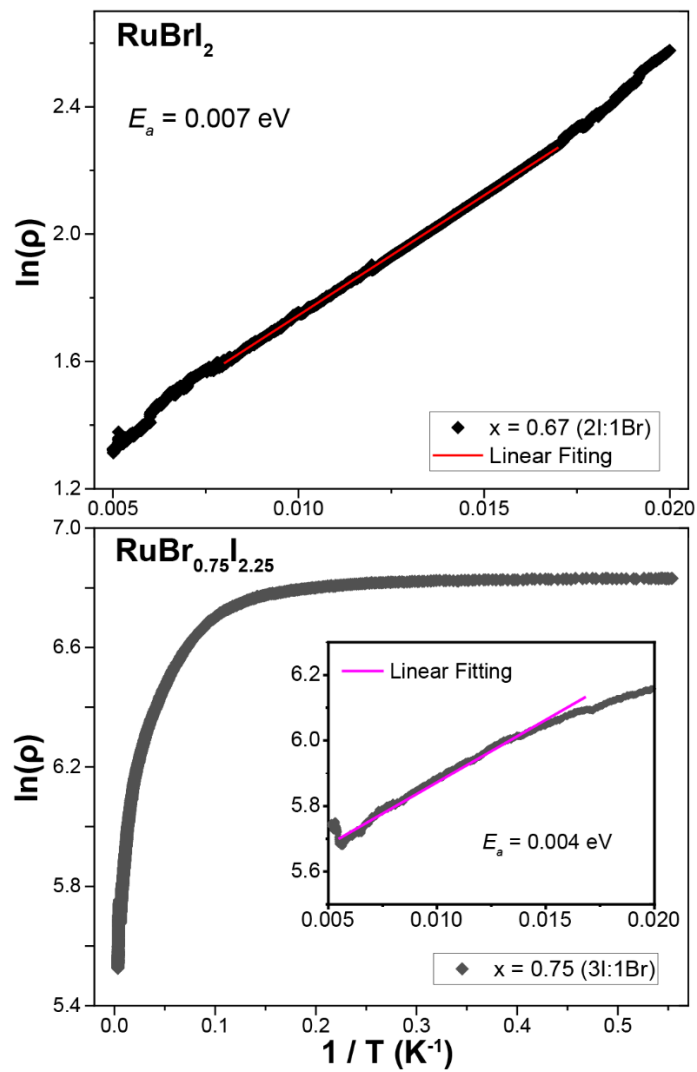

**Figure S5.** The plots of ln(ρ) vs 1/T of RuBrI$_2$ (top) and RuBr$_{0.75}$I$_{2.25}$ (bottom) samples. Linear fitting of a high temperature range is conducted to calculate the activation energy. The nonlinear shape of RuBr$_{0.75}$I$_{2.25}$ sample's plot suggests an increased metallic component with higher iodine content in this sample.

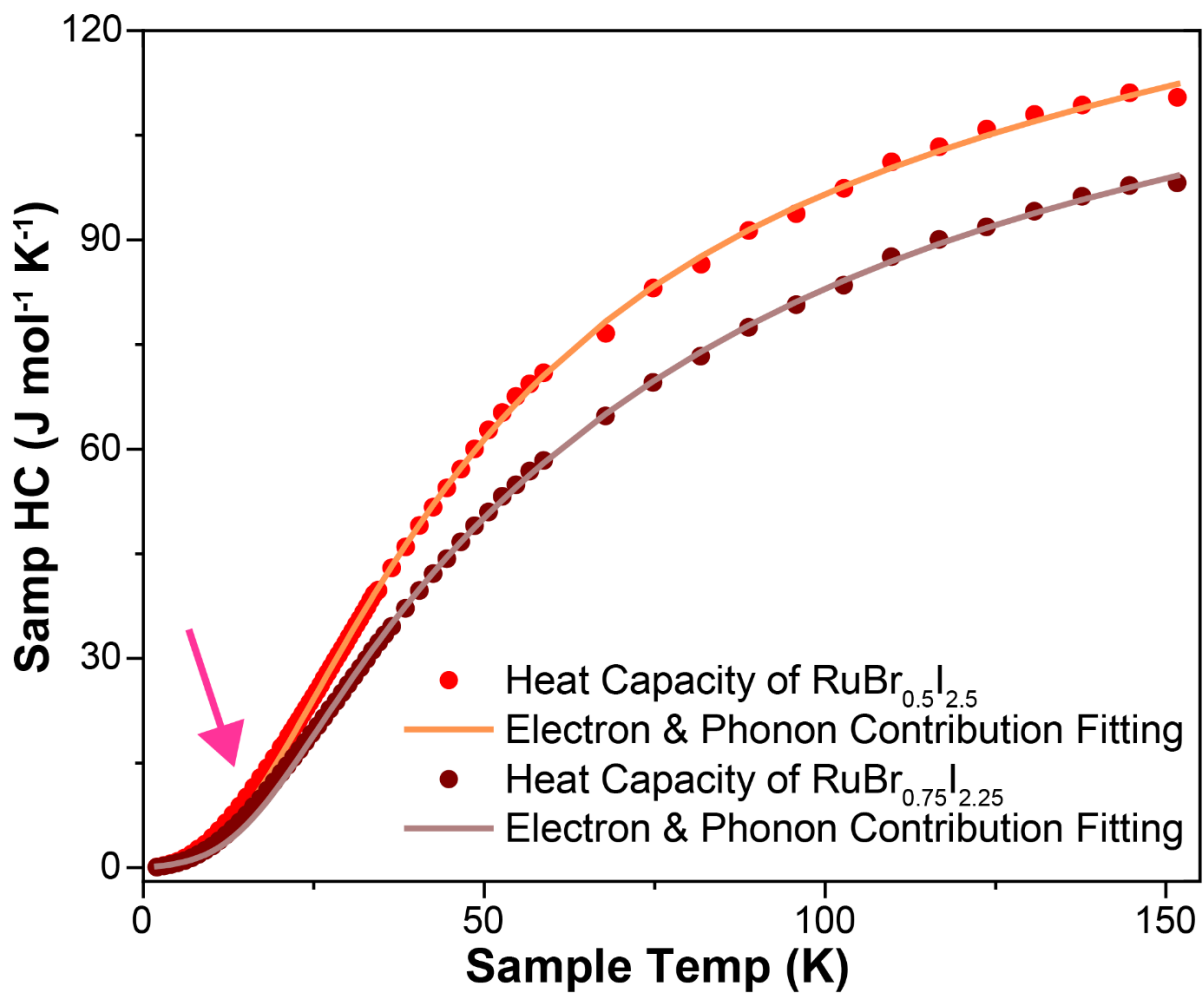

**Figure S6.** Heat capacity data of RuBr$_{0.5}$I$_{2.5}$ (red) and RuBr$_{0.75}$I$_{2.25}$ (brown) samples plotted versus temperature (2 to 150 K). The fitting of phonon and electron contribution is conducted for both plots using their high temperature range, and a small deviation between the data and the fitting can be observed at low temperatures (labeled with the arrow) suggesting the magnetic contribution of the systems.